\documentclass[10pt,letterpaper]{article}
\usepackage{opex3}
\usepackage{cite}
\usepackage{amsmath}

\begin{document}

\title{High-performance 3D waveguide architecture for astronomical pupil-remapping interferometry}

\author{Barnaby Norris\textsuperscript{1}, Nick Cvetojevic\textsuperscript{1,2,3}, Simon Gross\textsuperscript{2,4}, Nemanja Jovanovic\textsuperscript{5}, Paul N. Stewart\textsuperscript{1}, Ned Charles\textsuperscript{6}, Jon S. Lawrence\textsuperscript{3,4,7}, Michael J. Withford\textsuperscript{2,4,7}, Peter Tuthill\textsuperscript{1}}

\address{\textsuperscript{1}Sydney Institute for Astronomy, School of Physics, University of Sydney, NSW 2006, Australia; 
\textsuperscript{2}Centre for Ultrahigh Bandwidth Devices for Optical Systems (CUDOS), Australia; 
\textsuperscript{3}Australian Astronomical Observatory, NSW 2121, Australia; 
\textsuperscript{4}MQ Photonics Research Centre, Dept. of Physics and Astronomy, Macquarie University, NSW 2109, Australia; 
\textsuperscript{5}National Astronomical Observatory of Japan, Subaru Telescope, 650 N. A’Ohoku Place, Hilo, Hawaii 96720, U.S.A; 
\textsuperscript{6}Discipline of Medical Radiation Sciences, Faculty of Health Sciences, University of Sydney, Sydney, NSW 2006, Australia; 
\textsuperscript{7}Macquarie University Research Centre in Astronomy, Astrophysics \& Astrophotonics, Dept. of Physics and Astronomy, Macquarie University, NSW 2109, Australia 
}
\email{bnorris@physics.usyd.edu.au} 



\begin{abstract}
The detection and characterisation of extra-solar planets is a major theme driving modern astronomy, with the vast majority of such measurements being achieved by Doppler radial-velocity and transit observations.
Another technique -- direct imaging -- can access a parameter space that complements these methods, and paves the way for future technologies capable of detailed characterization of exoplanetary atmospheres and surfaces.
However achieving the required levels of performance with direct imaging, particularly from ground-based telescopes which must contend with the Earth's turbulent atmosphere, requires considerable sophistication in the instrument and detection strategy. 
Here we demonstrate a new generation of photonic pupil-remapping devices which build upon the interferometric framework developed for the {\it Dragonfly} instrument: a high contrast waveguide-based device which recovers robust complex visibility observables.
New generation Dragonfly devices overcome problems caused by interference from unguided light and low throughput, promising unprecedented on-sky performance. 
Closure phase measurement scatter of only $\sim 0.2^\circ$ has been achieved, with waveguide throughputs of $> 70\%$. 
This translates to a maximum contrast-ratio sensitivity (between the host star and its orbiting planet) at $1 \lambda/D$ (1$\sigma$ detection) of $5.3 \times 10^{-4}$ (when a conventional adaptive-optics (AO) system is used) or $1.8 \times 10^{-4}$ (for typical `extreme-AO' performance), improving even further when random error is minimised by averaging over multiple exposures. 
This is an order of magnitude beyond conventional pupil-segmenting interferometry techniques (such as aperture masking), allowing a previously inaccessible part of the star to planet contrast-separation parameter space to be explored.
\end{abstract}



\bibliographystyle{osajnl}


\section{Introduction}
\label{intro}
Ever since the beginning of the modern era of extra-solar planet discovery \cite{Mayor1995}, the detection and characterisation of exoplanets has been one of the most active areas in contemporary astronomy. 
Precise observations of exo-planetary systems promise to reveal the underlying physical mechanisms by which planetary systems - such as our own solar system - were formed, and estimate the ubiquity and diversity of earth-like planets in the galaxy. 
The vast majority of exoplanets detected thus far have been via techniques such as transits (wherein the light of the host star is observed to dim as the planet passes across it) or radial velocity (wherein the motion of the host star caused by the pull from its orbiting planets is detected via doppler shift) \cite{Wright2011}. 
While these techniques have been very successful in detecting and measuring a large number of exoplanets, they are limited to a restricted parameter space (for example, there are heavy observational biases favoring large planets and close orbits). 

Direct imaging of exoplanets -- wherein the star and nearby planet are separately resolved at an image plane -- stands to provide dramatic new insight into the origin and structure of exoplanetary systems. 
However the few exoplanets imaged thus far have been limited to wide apparent separations \cite{Marois2010} due to the challenging nature of high contrast measurement at very small spatial scales. 
Coronagraphs fed by AO systems represent the most developed class of high contrast imaging techniques, and although they have demonstrated exceptionally high contrast at large separation, performance is more limited at spatial scales of order $1 \lambda/D$ (corresponding to the Earth-Sun separation at a distance of $\sim$30 parsecs), even with the most advanced refinements \cite{Guyon2003}. 
To some extent this problem of the most productive search space lying within the so-called {\it inner working angle} of the coronagraph is inherent to the basic design of the instrument, and in practice is compounded by residual phase-aberrations present in the imaging system (largely from imperfect AO correction).

One solution to this inner working angle problem is aperture-masking \cite{Tuthill2000}, wherein the pupil of a large telescope is divided into a number of small sub-pupils using an opaque mask placed at the pupil plane, turning the telescope into a sparse interferometer array. 
Each pair of holes in the mask forms a baseline, and the key requirement is that the vector separation between any two holes in the mask is unique: such a mask is said to be \emph{non-redundant}. 
By analysis of the resulting interference pattern in the Fourier domain, phase-independent observables such as the squared visibility (the power spectrum of the image) and the closure phase (described in Section \ref{ClosurePhase}) can be derived. 
Since these observables are largely robust to residual wavefront phase aberration, the telescope's diffraction-limited performance can be recovered.

This technique has been successfully used to recover diffraction limited images at high contrasts, including the recent detection of sub-stellar companions undergoing the process of planetary formation \cite{Huelamo2011, Kraus2012} wherein contrast ratios of $\sim 300:1$ have been achieved. 
However, while this is a powerful tool for high-resolution imaging, its applicability to exoplanetary imaging is limited by several aspects of the experimental design.
Firstly, the requirement that the sub-aperture positions be non-redundant severely limits the fractional pupil area passed by the mask. 
For example, a commonly used 9-hole mask has a throughput of only $\sim 12 \%$, restricting the technique's use only to bright targets. 
A further limitation to the signal-to-noise ratio is imposed by the non-zero size of the sub-apertures and integration times. 
Closure phases are strictly immune to wavefront phase errors only in the limit of a point-sample of the wavefront in both space and time. 
In practice, neither is possible for the case of masking interferometry, which fundamentally limits the precision attainable by this technique.

\subsection{The Dragonfly instrument}
To address these limitations, the concept of a pupil-remapping interferometer was born \cite{Jovanovic2012, Huby2012}. 
Instead of using an aperture mask, the pupil of the telescope is divided into a number of segments, and each segment is injected into a single-mode fiber or waveguide. 
These then coherently remap the 2-dimensional pupil into a linear array, which then forms a 1-dimensional interference pattern. 
This accomplishes two things. 
First, while the output arrangement of fibers or waveguides needs to be non-redundant, the input arrangement does not, therefore allowing the entire pupil to be sampled, resulting in overall throughputs that at least in principle approach 100\%.
Alternatively, the 1-dimensional output array is now suitable to feed a lithographic photonic beam combiner \cite{Benisty2009}. 
Complete sampling of the pupil also provides much better Fourier coverage than an aperture mask. 
Second, since the light guides are single-moded, any phase-variation across a single `sub-aperture' is removed; it is spatially filtered. 
This means that the assumption of single phase for each sub-aperture is now valid, and the criterion for closure phases will strictly apply.

However, this technique introduces a new stringent requirement: since the light in all waveguides must remain coherent in order to form the interference pattern, the optical path-lengths of each of the fibers/waveguides must be precisely matched. 
For typical astronomical bandwidths ($\sim$ 50~nm at $\lambda = 1.5~\mu$m), this means path lengths must all be matched to within a few microns. 
For an optical fiber based remapper (such as in the FIRST instrument \cite{Huby2012}), this is a challenging tolerance since not only must the physical lengths of all fibers (and accompanying connectors) be precisely matched, but also any strain or temperature differences between the fibers must be carefully managed to avoid varying optical path lengths. 

The Dragonfly instrument uses an alternative technique -- the pupil is remapped using a monolithic photonic pupil-remapping chip. 
Here, a set of waveguides is inscribed into a single block of glass using the femtosecond laser direct-write technique \cite{Nolte2003, Gattass2008, Thomson2011}. 
By focusing a femtosecond laser into a block of glass (causing a local, permanent refractive index change), and translating the glass in three dimensnons, an arbitrary set of waveguide trajectories can be sculpted. 
The key advantage here is that since the routing of the waveguides can be precisely specified, path-length matching is relatively straightforward. 
Moreover, since the device is embedded within a single, monolithic block, differential strain or temperature changes between waveguides are eliminated. 
This photonic chip (referred to henceforth simply as the `pupil remapper') is integrated into the larger Dragonfly instrument, which provides beam handling, injection optimisation and detection. 
Further discussion of challenges and features of pupil remapper design is given in Section~\ref{EarlyDesignLimits}.

This technique was demonstrated on-sky with the Dragonfly instrument in 2011 \cite{Jovanovic2012}, and while this validated the photonic pupil remapper concept, levels of performance were insufficient to be competitive in astronomical research (particularly closure-phase precision and throughput). 
Subsequent refinements have produced a new generation of pupil remappers which address these concerns and provide performance levels approaching ideal.
These improvements, which promise a fully science-ready instrument in the next generation, are the subject of this paper.

\subsection{The Closure Phase observable}
\label{ClosurePhase}
The key data product delivered by Dragonfly is the \emph{closure phase} \cite{Baldwin1986}. 
This observable has been the key to successful high resolution, high contrast studies with conventional aperture-masking interferometry \cite{Lacour2011}, and becomes significantly more powerful when implemented with Dragonfly.

For an ideal imaging system, the observed phase of fringes from each baseline can be used to construct an image (or fit to a model.) 
However for astronomical imaging, the phase at each sub-aperture is randomised by the Earth's turbulent atmosphere. 
Even when adaptive-optics systems are employed, residual phase variation can be between 20$^\circ$ and 60$^\circ$ RMS, depending on conditions and wavelength. 
However, the closely related closure-phase is largely immune from these effects.

Consider a set of three sub-apertures (labelled 1, 2 and 3), forming three baselines (1-2, 2-3 and 3-1) in a closed triangle. Each sub-aperture is considered to have a random phase error from the atmosphere -- $\epsilon_1,~\epsilon_2$ and $\epsilon_3$. Since the absolute value of these errors is arbitrary, $\epsilon_1$ is set to zero, and so the phases measured on the three baselines are then
\begin{align}
    \label{eq:blPhases}
    \Psi_{\text{1-2}} = \phi_{\text{1-2}} + \epsilon_2 \nonumber \\
    \Psi_{\text{2-3}} = \phi_{\text{2-3}} + \epsilon_3 - \epsilon_2 \\
    \Psi_{\text{3-1}} = \phi_{\text{3-1}} - \epsilon_3 \nonumber
\end{align}
where $\phi_{ab}$ is the true phase of baseline $ab$ and $\Psi_{ab}$ is the phase measured on baseline $ab$. The closure phase $(CP)$ is then defined as the sum of these three baselines:
\begin{align}
    \label{eq:blSum}
    CP &= \Psi_{\text{1-2}} + \Psi_{\text{2-3}} + \Psi_{\text{3-1}} \nonumber \\
    &= (\phi_{\text{1-2}} + \epsilon_2) + (\phi_{\text{2-3}} + \epsilon_3 - \epsilon_2) + (\phi_{\text{3-1}} - \epsilon_3) \\
    &= \phi_{\text{1-2}} + \phi_{\text{2-3}} + \phi_{\text{3-1}} \nonumber
\end{align}
The atmospheric terms $\epsilon$ cancel and hence the closure phase is purely a function of the true phase. Interferometric observations usually consist of a large number of frames of fringes (each with short integration times). Instead of simply averaging the closure phase of each set of fringes, the \emph{bispectrum} is instead accumulated. The bispectrum is defined by the triple-product of the complex visibilities of each baseline, i.e.
\begin{align}
    \label{eq:triple}
    Bispectrum =  \widetilde{V_{\text{1-2}}} * \widetilde{V_{\text{2-3}}} * \widetilde{V_{\text{3-1}}}
\end{align}
The bispectrum of each frame is added together, and the argument of this sum is the closure phase. This method has the advantage that frames with low visibilities (e.g. due to particularly bad seeing at that time) contribute less to the final closure phase than frames with high visibilities.

The derivation above assumes that the phase error at each sub-aperture can be characterised by a single scalar phase term $\epsilon$. 
In the case of aperture masking this is only approximately true, limiting the precision of closure phase measurement. 
However when using a single-mode waveguide based remapper, such as in Dragonfly, the phase structure within each sub-aperture is filtered into a single mode with a single phase term, meaning the closure-phase assumption is now rigorously true, allowing more precise calibration and greater closure phase accuracy.

The closure phase precision directly relates to the contrast ratio obtainable at a given resolution. 
Simulations \cite{Lacour2011} have shown the following relationship between closure phase precision ($\sigma_{CP}$, in degrees) and contrast ratio achievable (at 1$\sigma$ detection), at a separation between star and planet at the telescope diffraction limit ($\lambda/D$):
\begin{align}
    \label{eq:contrastratio}
    \text{Contrast ratio detection}(1\sigma) = 2.5 \times 10^{-3} \times \sigma_{CP}
\end{align}
Therefore the observation of faint planetary companions is directly a function of the closure phase stability. 
This makes closure phase precision the primary figure-of-merit when discussing this type of interferometer.

\section{Experimental Setup}
\label{ExpSetup}
\begin{figure}[htbp]
\centering\includegraphics[width=1.0\textwidth]{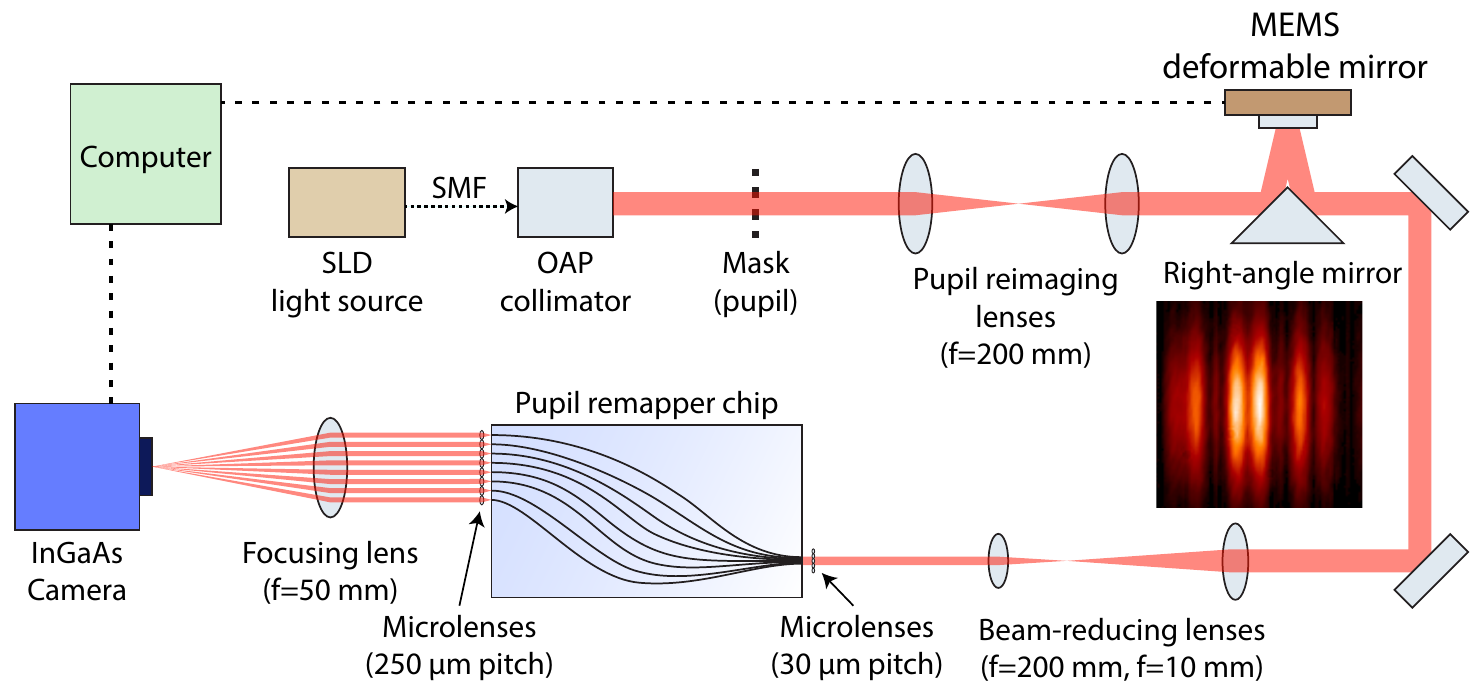}
\caption{Schematic diagram of the optical testbed used to produce interferometric data. See text for a full description. Inset: the fringe pattern produced on the detector. Three fringe frequencies are present, corresponding to the three baselines, which are extracted via Fourier transform.}
\label{fig_expsetup}
\end{figure}

The optical testbed which produced the measurements presented in this paper is shown in Figure~\ref{fig_expsetup}.
Light from a super-luminescent diode ($\lambda$ = 1550 nm, bandwidth FWHM = 50 nm) was propagated via a single-mode fiber to an off-axis parabolic collimator, from which the beam passes through the laser-cut aperture mask before being directed onto a MEMS segmented deformable mirror. The pupil-plane mask is reimaged onto the MEMS by the relay optics. 
The MEMS, manufactured by IrisAO, consists of 37 hexagonal segments, each of which can be precisely controlled in tip, tilt and piston (to a precision of $\sim$\,0.01\,milliradians). 
The mask ensures that only segments corresponding to the 8 waveguides in the prototype device (or fewer, if the experiment requires it) are illuminated. 
The reflected beam is then re-imaged by beam-reducing optics onto a hexagonal microlens array (MLA) with 30\,$\mu$m pitch, such that there is a one-to-one correspondence between MEMS mirror segments and individual MLA lenslets. 
This in turn is matched one-to-one with the injection points of the array of waveguides on the input-face of the photonic chip, such that each MLA lens injects the light from a single MEMS mirror segment into a single waveguide, with a matched numerical aperture. 
Light then propagates to the output end-face of the chip via the waveguides, whereupon each is re-collimated by another matched microlens array (with 250 $\mu$m pitch). 
The set of collimated beams are then all focused onto an infrared array detector (Xenics Xeva InGaAs camera) forming an interference pattern. 
The two microlens arrays and the remapper chip are each mounted on 5-axis translation stages to allow precise ($\sim$\,1\,$\mu$m) alignment, and injection for each waveguide individually optimised by steering the tip and tilt of the corresponding MEMS segment under computer control.
The setup used here was developed from that tested on-sky as described in 2011 \cite{Jovanovic2012}. 
 
The experimental aim is to measure the performance of the Dragonfly instrument, with the primary metric being closure phase precision, when subjected to phase errors typical of the Earth's turbulent atmosphere. 
First a set of three waveguides are selected which form a non-redundant array at the chip's output face (depicted in Figure~\ref{fig_chipendstrt}). 
For example, if the 1$^{\rm st}$, 3$^{\rm rd}$ and 7$^{\rm th}$ waveguides were chosen then this would form baselines of length 2, 4 and 6 units. 
Remaining waveguides are `switched off' by steering the appropriate MEMS mirror segment, preventing light from coupling at the waveguide input. 
The effect of atmospheric seeing was quantified by varying the input wavefront phase using piston introduced at the corresponding MEMS mirror segment, then recording the effect on the interference pattern (ideal performance would imply the closure phase remains constant regardless of phase errors introduced at the input). 
The quantitative behavior of closure phases extracted from each interference pattern, in particular their variation as a function of input wavefront phase error, allows the inherent measurement stability of the instrument to be measured and extrapolated to on-sky performance.

\begin{figure}[htbp]
\centering\includegraphics[width=1.0\textwidth]{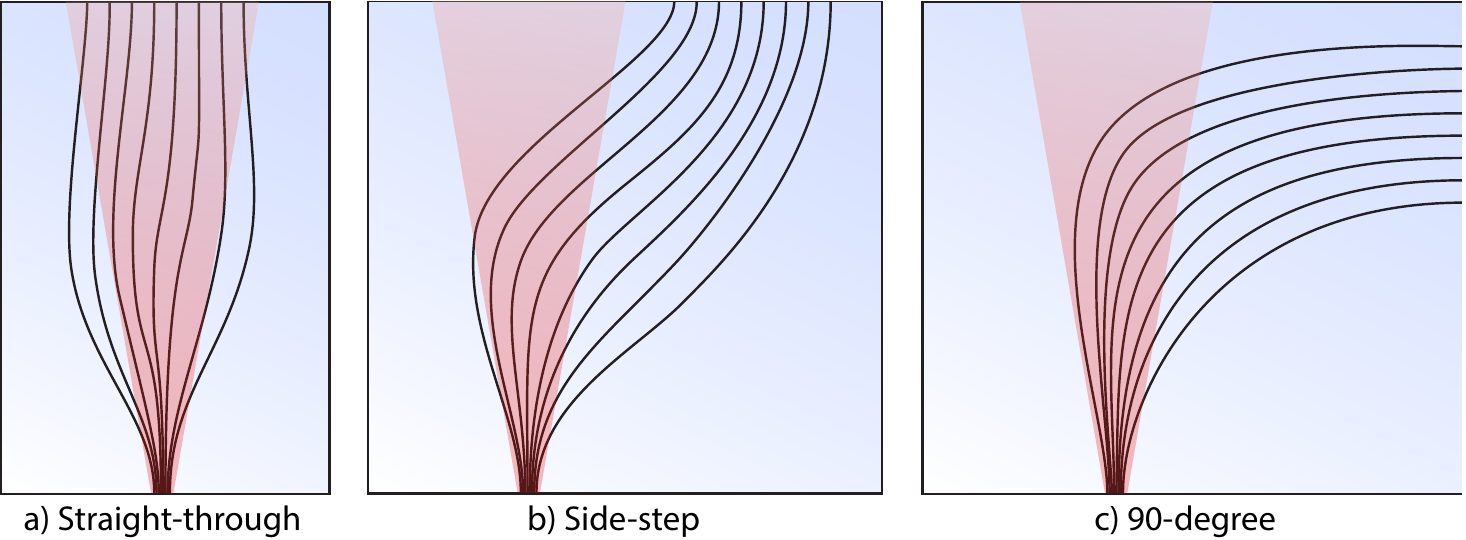}
\caption{Diagrams showing the three pupil-remapper topologies tested. 
The `side-step' and `90-degree' versions are designed to mitigate the interference effects of unguided light, by moving the waveguide outputs outside the cone of unguided stray light (shown in red). Additionally, topologies (a) and (b) have been manufactured and tested in two versions: the original design (denoted `old-generation') and the improved design (denoted `new-generation') which feature the advancements described in the text. 
The sketches given are illustrative only: actual waveguides are carefully designed (in three dimensions) to be of equal optical path-length.}
\label{fig_chipdiags}
\end{figure}

\section{Results: old-generation photonic pupil remappers}
\label{EarlyDesignLimits}
Using the measurement technique above, the closure-phase precision of old-generation pupil remapper chips was evaluated. 
While these early designs \cite{Jovanovic2012} were demonstrated on-sky to deliver the basic required functionality, their performance was limited. 
Poor performance was most clearly manifested as unstable closure-phase data. 
As described in Section \ref{ClosurePhase}, the performance metric we adopt here is the standard deviation of the closure phase of a baseline whilst the phase of one of its waveguides is pistoned through $2\pi$ radians (or multiple sets thereof), i.e. $\sigma_{\rm CP}$. 
This phase perturbation is worse than the wavefront error encountered on-sky with an adaptive optics system, where typical values around 20$^\circ$ to 60$^\circ$ RMS are encountered in the near-IR. 
The appropriate corrections for calculating the actual on-sky precision of Dragonfly are discussed in Section \ref{NewPerf}.

\begin{figure}[htbp]
        \centering
        \includegraphics[width=0.7\textwidth]{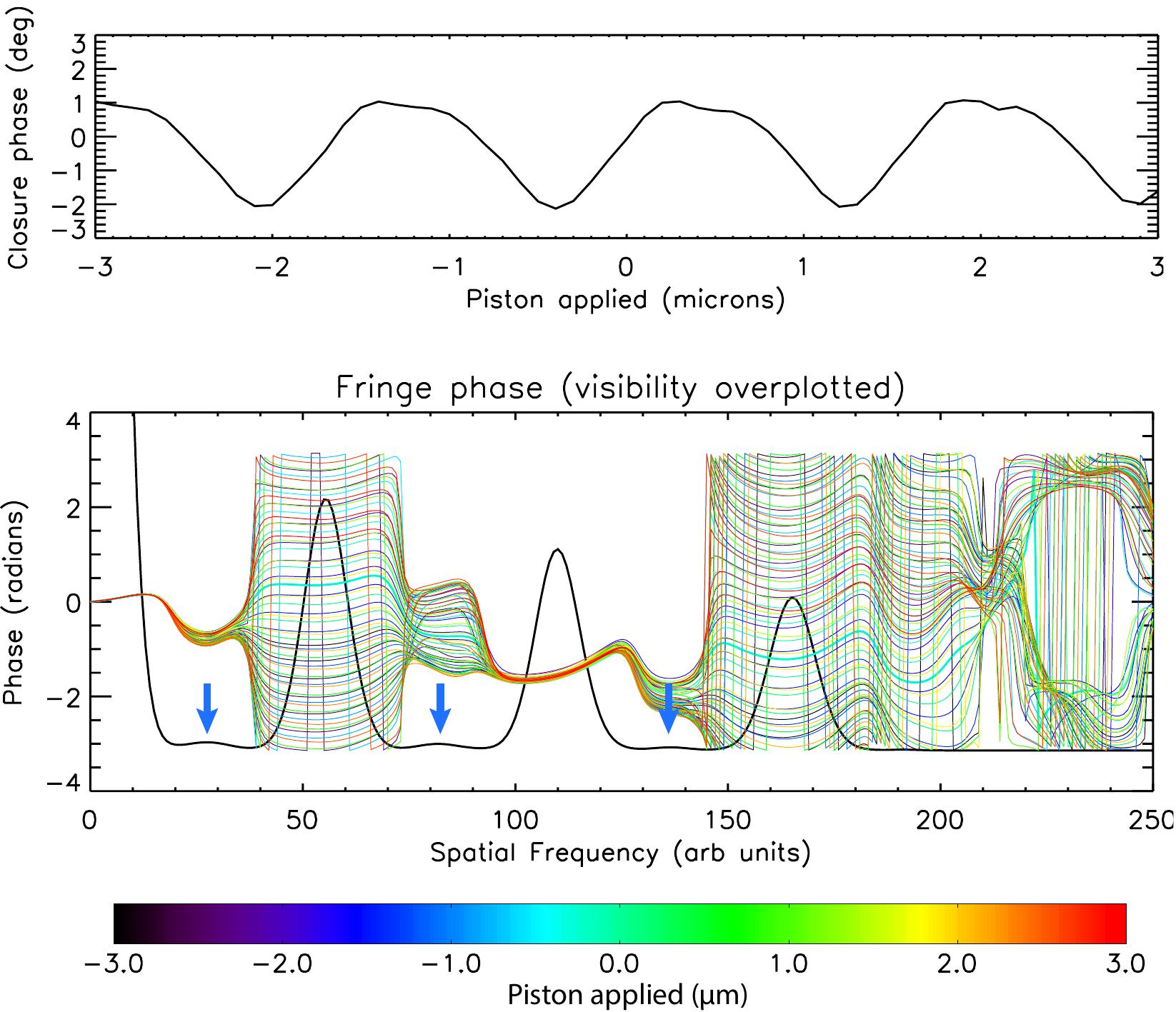}
        \caption{A set of closure phase measurements, recovered while adding successive increments to the piston in one waveguide.
Here, three waveguides for the original (straight-through) pupil remapper chip are illuminated. 
Top: the recovered closure phase as a function of applied piston. 
Ideally, the closure phase would remain constant, however a periodic variation of several degrees (of period 1$\lambda$) is seen. 
Bottom: The phase as a function of spatial frequency, where the applied piston offset is encoded in the colour bar. 
The power spectrum is overplotted in black (arbitrary units), showing three large peaks corresponding to the illuminated baselines. 
The phase is sampled at the spatial frequency corresponding to each of the peaks of the three waveguides; the vector sum of these three phases forms the closure phase. 
Small amounts of power are also seen between the expected peaks in the power spectrum (blue arrows) suggesting other waveguides are partially illuminated.}
        \label{fig_strtTypCPPlots}
\end{figure}

The goal was to consistently obtain $\sigma_{\rm CP}$ less than 1$^{\circ}$. 
While $\sigma_{\rm CP}$ as low as 0.4$^{\circ}$ were sometimes obtained with the original design, the performance was inconsistent and procedures such as optical realignment could result in large variations (as detailed in Section \ref{NewPerf}). 
Figure \ref{fig_strtTypCPPlots} gives an illustration of this problem with data obtained using the original pupil remapper in a configuration with three waveguides illuminated. 
In the top panel the closure phase is seen to vary periodically by several degrees as piston is added to one waveguide, with a period of 1$\lambda$ (ideal performance should remain constant). 
Furthermore, in the bottom panel some power is observed in the power spectrum (black line) outside of the three expected spatial frequencies (indicated by blue arrows), suggesting that ``switched-off'' waveguides have become illuminated, most likely by cross-coupling from the three guides in use.
Obtaining $\sigma_{\rm CP} < 1^\circ$ was found to be sensitively dependent on optical alignment, especially the precise positioning of the input and output microlens arrays with respect to the pupil-remapper chip. 
The limited performance encountered in early designs was found to be caused by two key problems: contamination from unguided `stray' light and high bend losses.

\subsection{Stray uncoupled light}
\label{StrayLight}
The coupling of light from the input microlens array into the waveguides is imperfect for several reasons.
The main ones are mismatch in numerical aperture, mode profile mismatch (the incoming beam is quasi-uniform, resulting in an Airy pattern at the focal plane, while the waveguide mode-field profile is Gaussian) and imperfect alignment of the microlens array. 
Measured coupling efficiency of a MEMS segment into a waveguide is between 60\% and 80\%. 
While in some applications the consequence of this would be limited to a mere loss in throughput, for coherent applications such as interferometry, the implications can be more serious.

\begin{figure}[htbp]
\centering\includegraphics[width=0.8\textwidth]{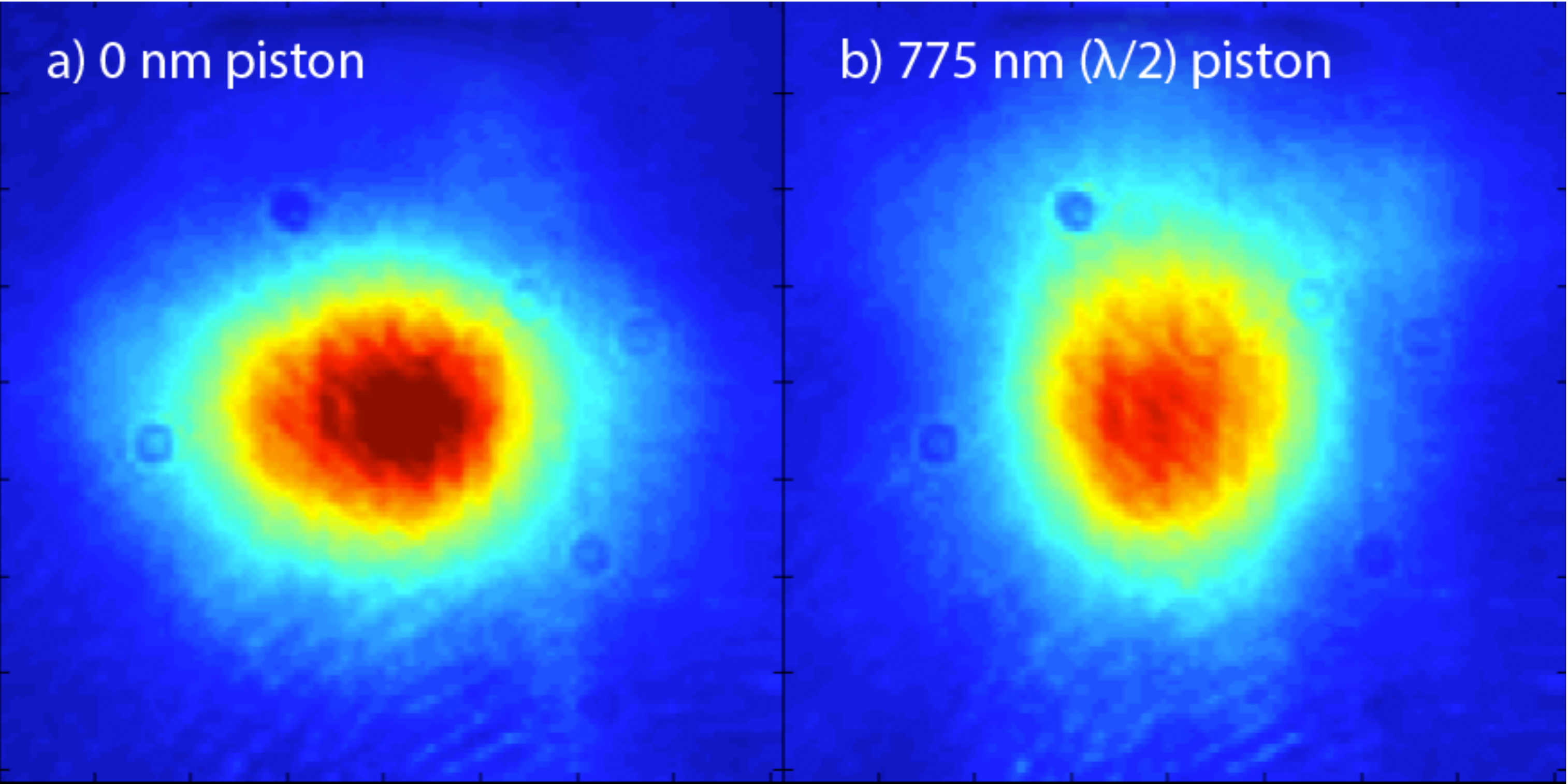}
\caption{The mode field at the output of a waveguide in the original photonic chip design, at $\lambda$ = 1550 nm, imaged with a 20X microscope objective. 
Panel (b) has a 775 nm ($\lambda / 2$) piston added to it (using the MEMS mirror) with respect to panel (a). 
The mode field is seen to deform, due to interference with coherent unguided background light. 
The background light itself is not visible at this camera exposure level.}
\label{fig_modeFieldsDeform}
\end{figure}

\begin{figure}[htbp]
        \centering
        \includegraphics[width=0.8\textwidth]{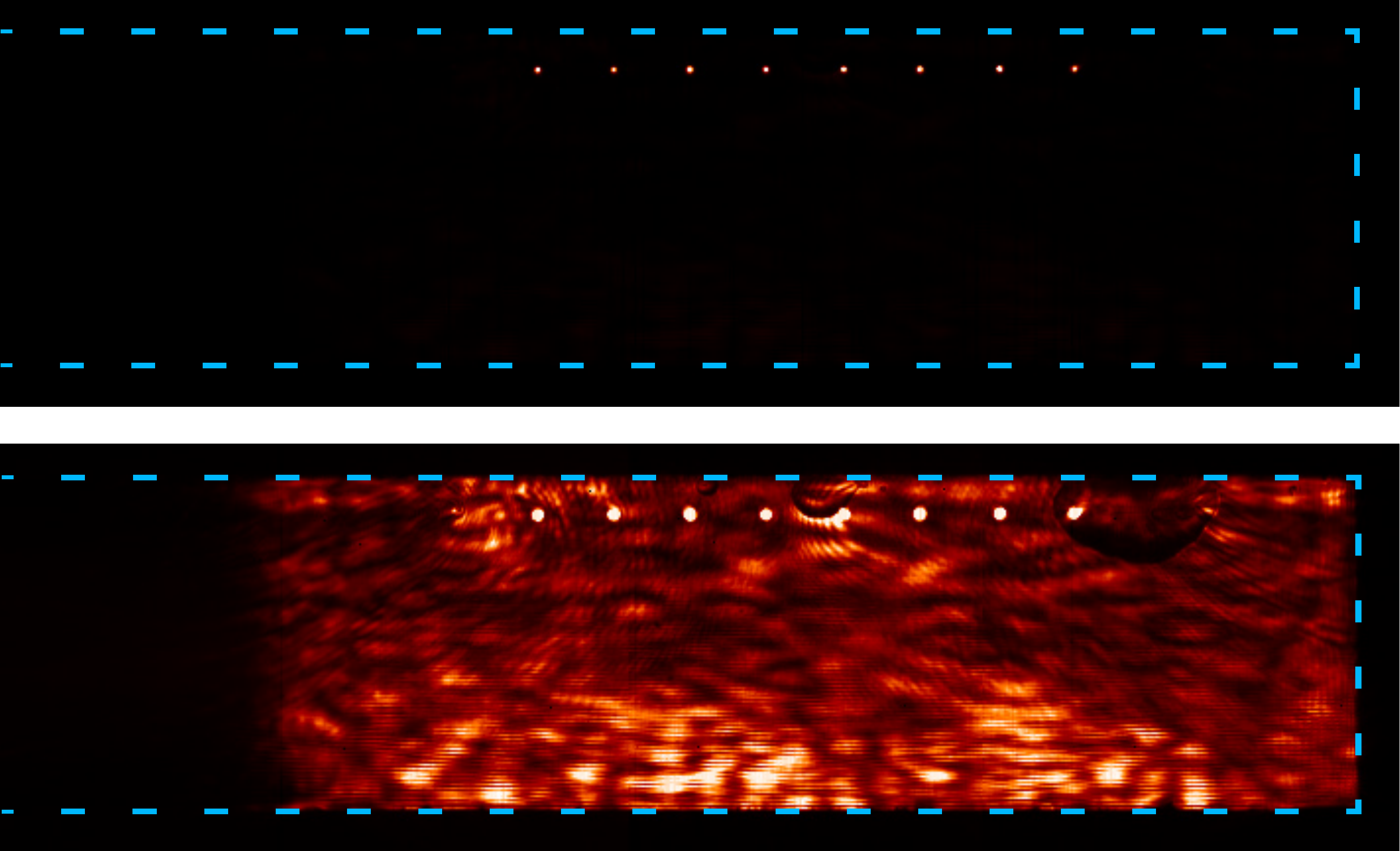}
        \caption{The output end-face of the original photonic chip, imaged with a 4X microscope objective at two exposures: top panel is 100 $\mu$s (chosen such that the waveguides do not saturate), bottom panel: 6400 $\mu$s. The dashed line indicates the edges of the chip.
All 8 waveguides are illuminated, while in the long exposure image the unguided background light is visible, adorned with fringes due to interference between various radiation fields. 
As the piston applied to any given waveguide is varied, the phase of the fringes adjacent to that waveguide's output is observed to shift proportionally. 
The decrease in background light toward the left-hand side of the image is due to the edge of the cone of unguided stray light projected from the input face.}
\label{fig_chipendstrt}
\end{figure}

The problem arises that stray un-coupled light propagates, unguided, through the bulk of the photonic device and interferes with the mode field at the waveguide outputs. 
A dramatic example of this is shown in Figure \ref{fig_modeFieldsDeform}, wherein the shape of the mode field at the waveguide output is seen to deform as the piston term of this waveguide is varied by half a wavelength. 
At this camera exposure level, while the unguided background light itself is not visible, its effect on the single-mode output waveguide profile, which distorts as a function of piston, is readily apparent.
The same exposure level is used for the wider view in Figure \ref{fig_chipendstrt} (top), wherein all 8 waveguides are visible. 
However when the exposure is increased by a factor of 64 in Figure \ref{fig_chipendstrt} (bottom), the background stray light is clearly visible as a complex fringe pattern. 

\begin{figure}[htbp]
\centering\includegraphics[width=1.0\textwidth]{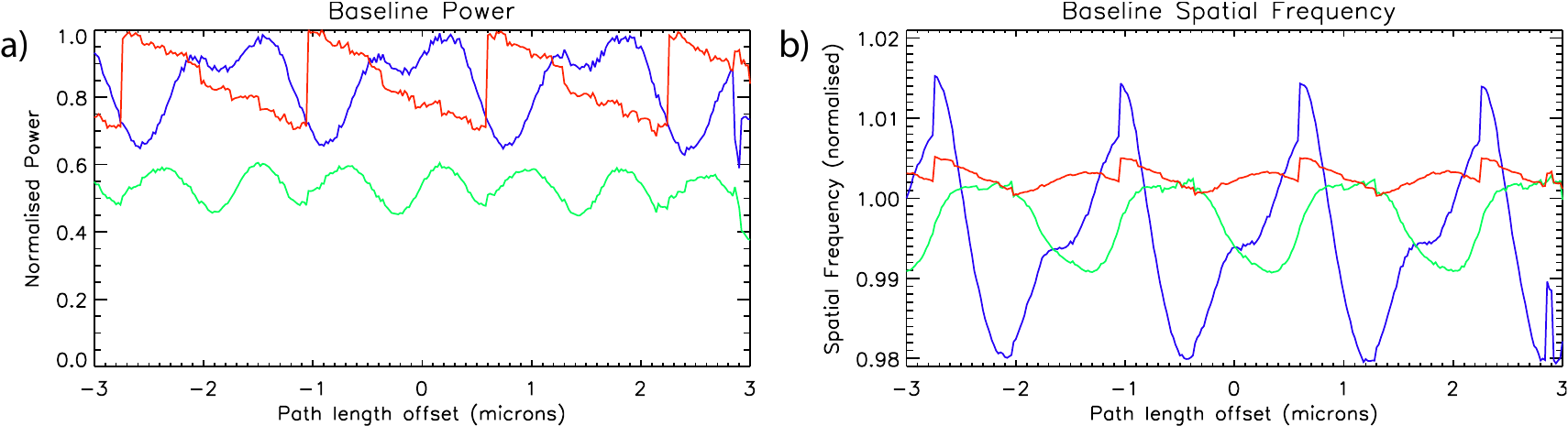}
\caption{The power (a) and spatial frequencies (b) of the three power spectrum peaks for a case where three waveguides are illuminated, and piston offset is varied. 
The three colours correspond to the three baselines. 
The spatial frequency has been normalized with respect to the expected spatial frequency for that baseline. 
As piston is varied, both the power and the spatial frequency of the baseline is seen to vary periodically. }
\label{fig_strt-peakposns}
\end{figure}

The effects of interference with stray light are quantified in Figure \ref{fig_strt-peakposns}. 
As the input waveguide phase offset is smoothly ramped, both the power in each baseline and the spatial frequency of the baseline is seen to vary periodically. 
The latter point is especially surprising since the spatial frequency corresponds directly to the baseline length -- that is, the physical separation of the waveguides at the output face of the photonic chip. 
However this can be understood when the deformation of the waveguide's mode field profile (Figure~\ref{fig_modeFieldsDeform}) is taken into consideration. 
The end result of interference caused by unguided stray light is to violate the fundamental assumption underlying closure phase -- that each of the three baselines in a closed triangle yields a single defined baseline length and a single phase, with no phase structure within a sub-aperture.
Stray light, in short, directly causes closure-phase measurement error.

\subsection{Waveguide bend losses}
\label{BendLoss}
Waveguides in the old-generation pupil remapper chips suffered from high bend losses \cite{Charles2012}, and hence required large bend radii in order to maintain useable throughputs. 
The waveguides in the original design discussed thus far had throughputs of between 54\% and 66\% (not counting coupling loss), which is somewhat lower than the 82\% throughput expected purely from loss due to absorption in the Eagle 2000 substrate \cite{Meany2014}. 
The low throughput problem was compounded by the strong effect of excess unguided light on closure-phase precision (Section \ref{StrayLight}).
Losses increase rapidly with decreasing bend radii. 
Bend-loss from a tight bend with ROC = 20~mm is 2.9 times higher than that of a wider bend with 40~mm ROC.
As will be detailed in Section \ref{NextgenRemappers}, an attempt to combat the problem of stray-light interference was made by positioning the waveguide's outputs outside of the cone of unguided light, using a `side-step' design (see Figure \ref{fig_chipdiags}b). 
However the decreased bend radii required to reroute the guides along a more tortuous path led to unacceptable bend losses, with throughputs of these waveguides being as low as 0.6\%.
Beyond the obvious undesirability of low throughputs in astronomical applications, the high bend losses had two additional effects which impacted negatively on closure phase stability. 

Firstly, light lost in bends contributed to the overall amount of unguided stray light within the chip, compounding the severity of the interference problem described in the previous section. 
This was exacerbated by the fact that light lost from waveguide-bends located near the output end of the chip has propagated along a similar optical path-length to -- and hence has a high degree of coherence with -- the guided light, resulting in stronger undesired interference at the output.

Secondly, light lost at bends may recouple into adjacent waveguides, causing cross-coupling. The average cross-coupled power between a pair of waveguides in the original remapper chip was of order $10^{-5}$, while in the side-step design the cross-coupled power ranged from $\sim 10^{-4}$ to as high as $6 \times 10^{-3}$. 

Cross-coupling of this magnitude causes closure phase error of order 1$^\circ$.
Inspection of the power spectrum in Figure \ref{fig_strtTypCPPlots} reveals small peaks at spatial frequencies in between the expected peaks, located at integer multiples of the unit baseline. 
This is consistent with dark waveguides being excited by cross-coupling with illuminated waveguides. 
This erodes the non-redundant property of the output array, with small amounts of power (and associated phase components) from diverse optical paths being blended into the measured baseline phases, again violating the closure-phase conditions.
 
A further limitation imposed by high bend loss is that it strongly limits the number of waveguides that can be incorporated into a device. 
The more waveguides incorporated into a device, the more complex their routing needs to be in order to maintain path-length matching and avoid clashes between adjacent tracks. 
For such designs to be feasible, much better optical performance at small bend radii than in the original 8-waveguide straight-through design (which is already at the limit of acceptable bend loss) was required. 
Ultimately the goal is to remap the entire telescope pupil into a guided structure, which will require several tens of waveguides (e.g. 37 with the current MEMS mirror).

\section{Results: new-generation photonic pupil remappers}
\label{NextgenRemappers} 

\subsection{Eliminating interference from unguided light} 
Light which is focused onto a waveguide by an individual microlens, but which does not couple into the waveguide, continues to propagate through the glass substrate in a cone corresponding to the numerical aperture of the focusing microlens -- see Figure \ref{fig_chipdiags}a. 
This light interferes with the guided light from the waveguides, compromising closure phase measurements as previously described. 
A solution is to move the outputs of the waveguides outside of this cone of unguided stray light, as illustrated in Figure \ref{fig_chipdiags}b, so that it no longer enters the downstream optics. 
This technique was previously attempted, however the small bend radii required to execute the sideways step led to extremely high bend losses, with waveguide throughputs as low as 0.6\% and not viable for astronomical science. 

These bend losses have been essentially eliminated with the new generation of chips fabricated using a thermal annealing process to optimise the refractive index profile of the waveguides \cite{Arriola2013}. 
In this refinement to the original direct-write technique, the original waveguides are inscribed with higher pulse energy, resulting in larger core guides (multimode at our wavelength). 
The device is then subject to a thermal annealing process, wherein the device is raised to a temperature above the substrate's annealing point, (but below its softening point) and then cooled adiabatically, both at precisely controlled rates. 
This has the effect of washing out a fraction of the refractive index modification in the originally inscribed waveguides, in particular removing unwanted structures in the periphery and leaving behind a single-mode waveguide with an optimised refractive index profile. 
Such annealed waveguides are now highly resistant to bend loss for two main reasons. 
Firstly, the core region of the waveguide has a higher refractive-index contrast due to the higher pulse energies during inscription. 
Secondly, the index profile of the resulting waveguide consists only of a Gaussian-like core, a geometry known to exhibit low losses during bends \cite{Arriola2013}.

The annealing technique allowed the production of remapper chips with routing designed to avoid the impact of unguided light, such as the `side-step' design mentioned above, while still maintaining excellent throughput. 
Additionally, a `90 degree bend' design was created, which placed the inputs and outputs of the waveguides on adjacent orthogonal faces of the chip - see Figure \ref{fig_chipdiags}c. 

\begin{figure}[htbp]
\centering\includegraphics[width=1.0\textwidth]{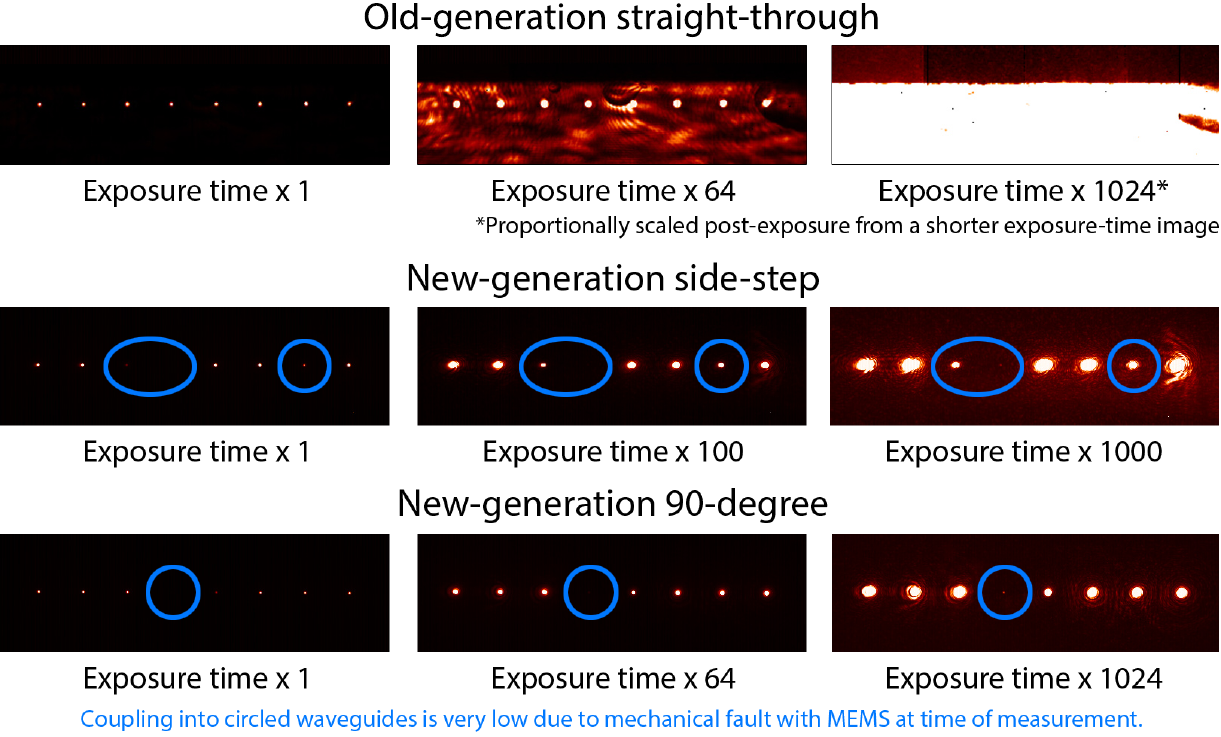}
\caption{Images of the output face of three chip designs, at various camera exposure times. The exposure time for the first (left-most) image was set such that the waveguides themselves just saturate. It is seen that the new-generation chips exhibit far less unguided light at the output face. Note that at the exposure levels needed to see any stray light for the new chips (right panels), the background light levels for the original chip saturate the detector.}
\label{fig_all-chipend-pics}
\end{figure}

As seen in Figure \ref{fig_all-chipend-pics}, the amount of unguided light visible at the end-face of the chip, together with the error due to stray light interference, was greatly reduced in the new chips.

\subsection{Optical performance of new-generation remapper chips}
\begin{figure}[htbp]
        \centering
        \includegraphics[width=\textwidth]{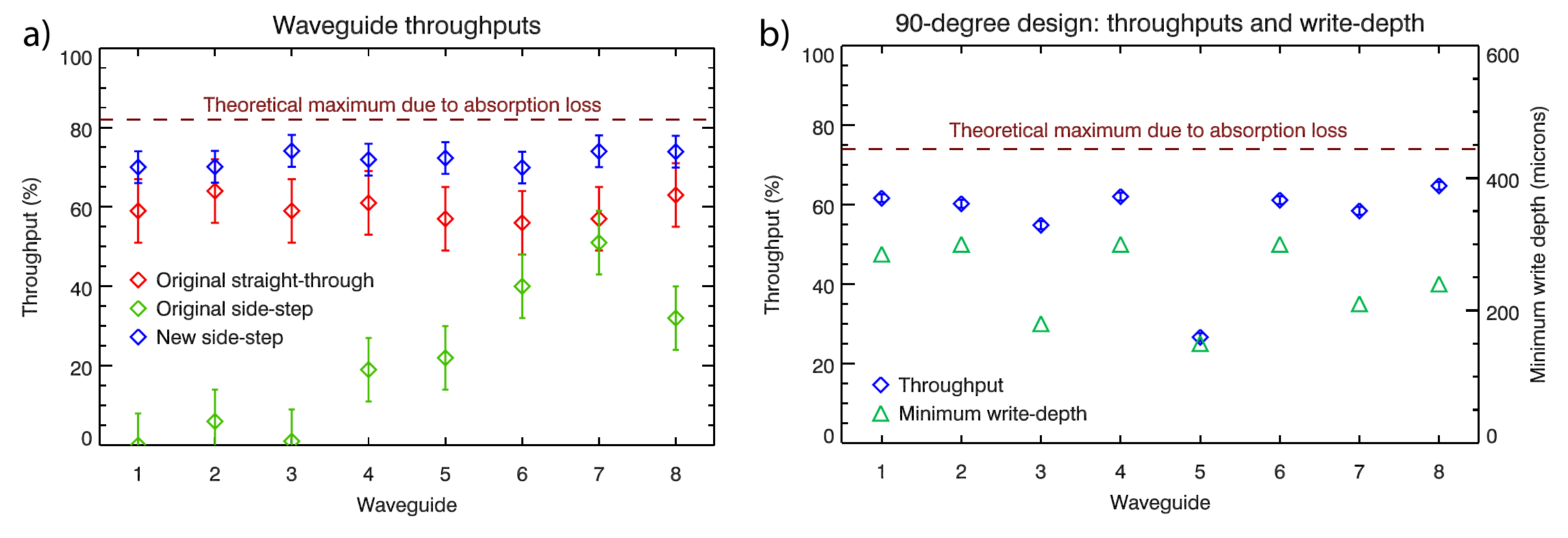}
        \caption{Waveguide throughputs for both the old- and new-generation remapper chips. 
Panel (a) gives throughputs for the original straight-through and side-step designs, and for the new side-step. 
Note that the throughputs for the new \emph{side-step} chip exceed those of the old \emph{straight-through} chip. 
Panel (b) gives the throughputs for the new 90 degree chip, along with minimum write depth. 
Waveguides with small write depths (i.e. guides are written close to the glass surface) suffer from poor throughputs.}
        \label{fig_throughputs}
\end{figure}

The throughputs for the new devices, along with those of the original device, are given in Figure \ref{fig_throughputs}. 
In panel (a) it is seen that while the old-generation side-step design suffered from very low throughputs, the new-generation side-step chip has throughputs approaching the maximum possible values set by material absorption. 
Moreover, the throughputs for the new-generation \emph{side-step} chip exceed those of the old-generation \emph{straight-through} chip. 
This design provides mitigation of the aforementioned unguided-light issue while maintaining high throughputs across all waveguides. 
Waveguides in the annealed chips show negligible bend-losses for radii of curvature as tight as 20\,mm \cite{Arriola2013}.

In panel (b) of Figure \ref{fig_throughputs}, the throughputs of the new-generation `90-degree' design are given, along with each waveguide's minimum write-depth -- that is, the minimum distance between the surface of the glass and the waveguide. 
While most waveguides perform well in terms of throughput, waveguides with smaller minimum write-depths are seen to have lower throughputs. 
This is due to problems with the oil-immersion objective lens and the high average laser powers ($\sim$500~mW) used during the fabrication process. The high average power causes a thermal lens within the oil layer resulting in defocusing of the laser beam. The thicker oil layer between objective and sample for low writing depth results in a stronger thermal lens and thereby causes a stronger distortion of the writing laser beam which impairs the waveguide quality.
Future generations of 90-degree chips will avoid this by using a more conservative minimum write-depth in their design.

New-generation chips also exhibited superior cross-coupling properties. 
The old-generation straight-through chip cross-coupling between waveguides was estimated\footnote{A more precise value could not be measured due to stray light contamination.} to be $\sim 10^{-5}$ while for the old-generation side-step it ranged from $\sim 10^{-4}$ up to $6 \times 10^{-3}$. 
However the cross-coupling for the new-generation side-step design was below the measurement threshold of $\sim2.5 \times 10^{-6}$ for 80\% of measurements, the exception being six waveguide-pairs\footnote{Here, we define a waveguide pair to be a given input waveguide and a given output waveguide, resulting in 56 pairs for the 8 waveguide chip} which show cross coupling ranging between $2.8 \times 10^{-6}$ and $1.2 \times 10^{-5}$. 

Cross-coupling in the new-generation 90-degree chip was found to be higher. 
The median cross-coupling was less than $4 \times 10^{-6}$ (most waveguide pairs being below the detection limit) however several pairs exhibited high cross-coupling, with 13 of the 56 pairs having cross-coupling $> 1 \times 10^{-5}$ and 3 pairs (7 $\rightarrow$ 8, 8 $\rightarrow$ 7 and 5 $\rightarrow$ 3) having cross-coupling $> 1 \times 10^{-4}$. 
Waveguide 5, seen in Figure~\ref{fig_throughputs} to have low throughput, exhibits the worst cross-coupling. 
Thus the new-generation side-step performs better than the old-generation straight-through design while mitigating stray light, although the new-generation 90-degree chip performs relatively poorly.

\subsection{Interferometric performance of Dragonfly with new-generation pupil remapper chips}
\label{NewPerf}
\begin{figure}[htbp]
        \centering
        \includegraphics[width=0.7\textwidth]{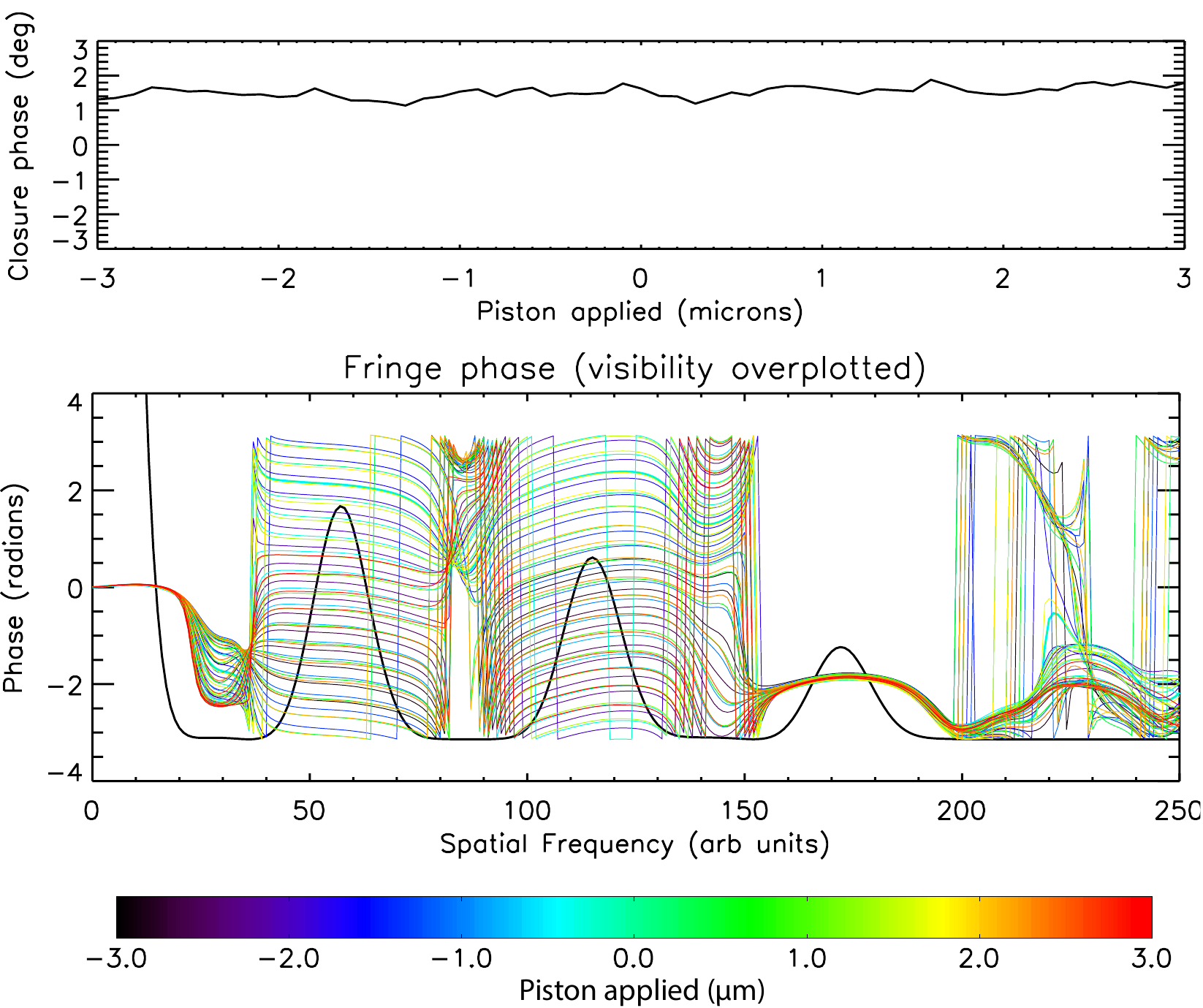}
        \caption{A set of closure phase measurements taken while successively incrementing piston in one waveguide for the new-generation side-step design pupil remapper chip. 
Here, three waveguides are illuminated forming a single closing triangle. 
Top: the closure phase is far more consistent as piston is applied (as compared with the old-generation chips shown in Figure \ref{fig_strtTypCPPlots}), with $\sigma_{\rm CP} = 0.22^\circ$. 
Bottom: the phase as a function of spatial frequency, for each piston offset (colours). 
The power spectrum is overplotted in black (arbitrary units). 
In contrast to the old-generation chip, only the three expected peaks in the power spectrum are seen.}
        \label{fig_ssCPPlots}
\end{figure}

The design goal of the pupil-remapper is to enable the Dragonfly instrument to consistently obtain closure-phase precisions (denoted $\sigma_{\rm CP}$) of less than $1^\circ$ when input phase errors $> 2\pi$ radians are applied. 
An example of the closure-phase results of the new-generation of side-step chips is shown in Figure \ref{fig_ssCPPlots}. 
Excellent closure phase stability is now recorded, with $\sigma_{\rm CP} = 0.22^\circ$ achieved while input wavefronts are pistoned through multiples of $2\pi$ radians, in a single measurement set.

Furthermore, the performance of the new-generation remappers is also far more robust and reproducible, no longer sensitive to small misalignments of the microlenses and the photonic chip. 
This is a key requirement, since deployment to an instrument platform on a telescope is dependent on fast alignment in hard-to-access spaces, tolerance of vibration and possibly a moving gravity vector if the instrument is mounted to the telescope itself (such as at a Cassegrain focus). 
To test this, the microlenses and photonic chip were deliberately misaligned, and then realigned and the closure-phase tests performed again. 
This cycle was repeated multiple times for each of the different chip designs. 
Each realignment took less than 1~minute and involved manipulating the translation of the chip and microlenses in X, Y and Z, and the roll angle (i.e. rotation about the axis parallel to the direction of propagation) of the chip. 
The goal in each realignment was simply to produce fringes on the detector, which have power visible in all three baselines of a triangle (indicating all three waveguides in the triangle are illuminated) and where the fringes appear parallel (eliminating any rotational misalignment). 

\begin{figure}[htbp]
\centering\includegraphics[width=0.7\textwidth]{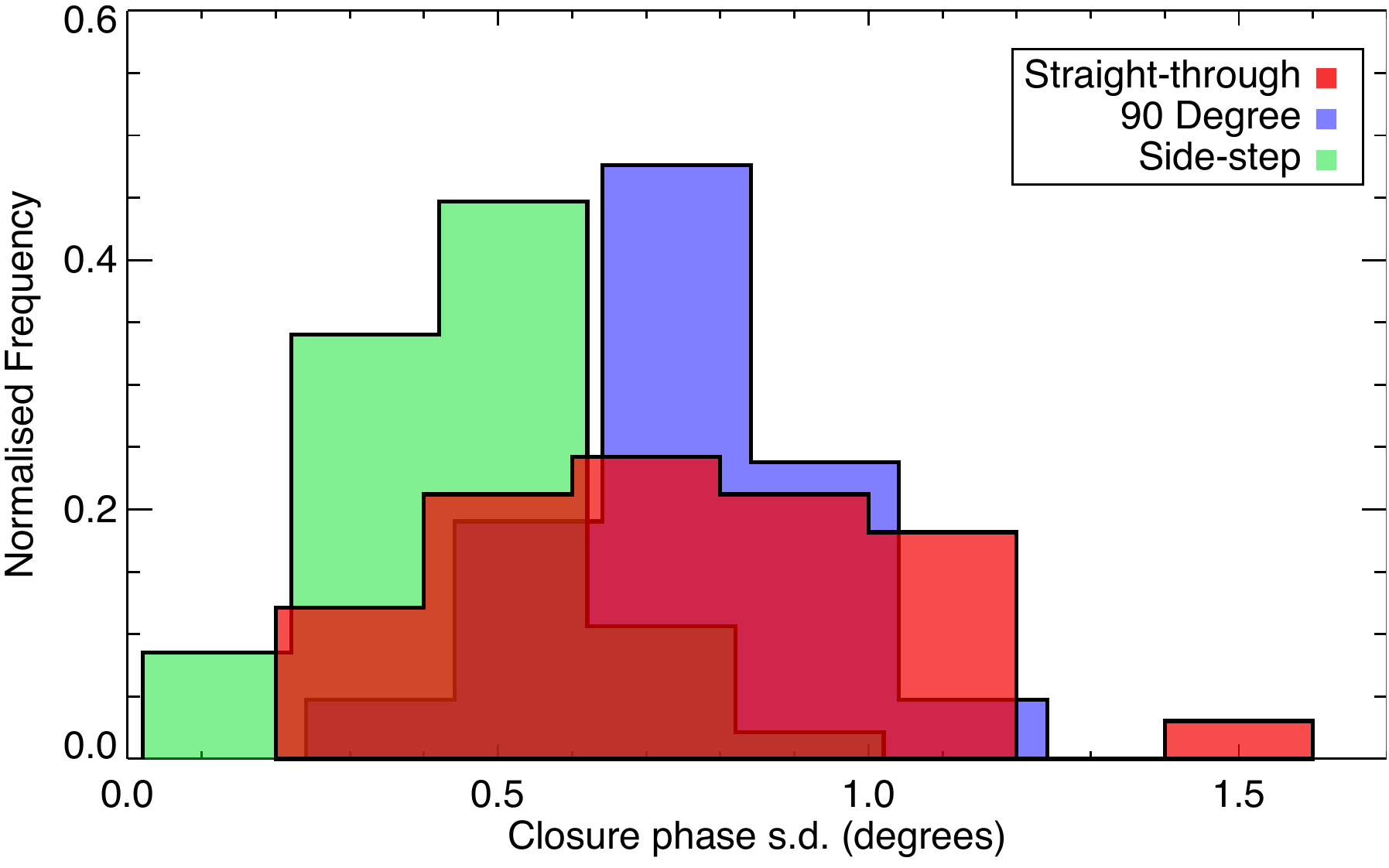}
\caption{A histogram showing the distribution of $\sigma_{\rm CP}$ when the pupil-remapper and microlens-array are subjected to repeated realignments. The frequencies have been normalized such that the integral of each histogram is unity.}
\label{fig_CPHist}
\end{figure}

A histogram of the test results is shown in Figure \ref{fig_CPHist}. 
While occasionally good performance is seen from the old-generation (straight-through) chip, this is not reliably reproducible and is highly sensitive to each realignment, with $\sigma_{\rm CP}$s ranging from 0.4$^\circ$ to 1.5$^\circ$. 
On the other hand, the new-generation side-step chip performs far better, with a median $\sigma_{\rm CP}$ of $0.42^\circ$. 
Moreover, this chip exhibited this performance consistently, with $\sigma_{\rm CP}$s better than $0.7^\circ$ in all but one measurement, and as low as $0.15^\circ$. 
The new-generation 90-degree design does not perform as well, with a median $\sigma_{\rm CP}$ of $0.74^\circ$. 
This is consistent with stray light arising from bend losses within the chip (due to its tighter bend radii than the side-step design) contaminating the measurements.

Correction for detector non-linearity was also important. The detector exhibited relatively little non-linearity, with $R^2 = 0.993$, with the most non-linear region being at the `toe' of the response function (where counts are $< 1000$ ADU). However this still had a large impact on the closure phase precision. When the non-linear correction is applied (derived from a polynomial fit to the measured detector response) the side-step chip exhibited $\sigma_{\rm CP} \approx 0.2 ^\circ$, but these same data yield $\sigma_{\rm CP} \approx 2.4 ^\circ$ when non-linear correction is neglected. This suggests careful non-linear correction should be applied whenever closure phases are measured from such interferograms, e.g. in aperture masking interferometry.

To convert these experimental results to real-world performance, a correction must be made for the fact that $2\pi$ radian phase error applied is worse than the actual phase error encountered behind an adaptive optics system, which is the intended platform for Dragonfly (use without an AO system is not feasible because the tip/tilt errors in the wavefront greatly reduce the efficiency of the coupling into the waveguides, leading to very low throughput). 
The phase errors applied were uniformly distributed between 0 and $2\pi$ radians, so the RMS error is 1.8 radians. 
On the other hand, the phase error was only applied to one of the three sub-apertures at a time, whereas in on-sky use all three sub-apertures would be subject to the error, so the resulting closure-phase error is $\sqrt{3}$ times worse. 
So, if the predicted AO-corrected phase error encountered is $\epsilon_{AO}$ (in radians), then the predicted closure-phase error is 
\begin{equation}
    \sigma_{CP}^{On-sky} = \sqrt{3} \cdot \frac{\epsilon_{AO}}{1.8} \cdot \sigma_{CP}^{Expt} 
\end{equation}
where $\sigma_{CP}^{Expt}$ is the experimental $\sigma_{CP}$ measured in these tests and $\sigma_{CP}^{On-sky}$ is that predicted on-sky. 
This assumes the closure phase error is linear with phase, a good approximation when the closure phase error is small as it is here \cite{Martinache2010}.

For a standard adaptive optics system, the residual RMS wavefront error is $\sim$250 nm, while for the new-generation extreme-AO systems this is as low as 80 nm \cite{Jovanovic2012}.
Thus the previously measured performance metric $\sigma_{\rm CP}^{Expt}$ of $0.22^\circ$ (from the new-generation side-step design) translates to 0.21$^\circ$ for a conventional AO system and $0.07^\circ$ for an extreme-AO system. 
This results in a contrast-ratio sensitivity limit at $1 \lambda/D$ (1$\sigma$ detection) of $5.3 \times 10^{-4}$ and $1.8 \times 10^{-4}$ respectively (see Equation  \ref{eq:contrastratio}). 
If instead the median performance of this chip of 0.42$^\circ$ is considered, then the $1 \lambda/D$ contrast ratio detectable is $1.0 \times 10^{-3}$ and $3.3 \times 10^{-4}$ respectively. This is well within the performance range required to directly detect young planets in star-forming regions (e.g. \cite{Kraus2012}).

However these calculations are extremely conservative. 
The $\sigma_{\rm CP}$ values quoted here refer to the standard deviation of the closure phase for a single measurement set (around 100 ms total integration time). 
In practice, the astronomical source would be observed for minutes or hours and a large statistical sample of measurements taken. 
The resulting closure phase would be the mean of the closure phases and the uncertainty would be the standard error in the mean. 
Ideally (in the absence of photon noise and systematic errors that are unable to be removed by observing a point-spread-function calibrator star) this error would go as $1/\sqrt{N}$ (where N is the number of measurements), so a 1~hour observation, consisting of $\sim$30~000 such measurements, would have a best-case closure phase precision of 0.22$^\circ / \sqrt{30000} = 10^{-3~\circ}$, or a contrast ratio detection limit of $\sim 2 \times 10^{-6}$.  
This level of performance puts the instrument in reach of the ultimate goal of imaging mature planetary systems \cite{Burrows2005}. 
This is a theoretical limiting best case; in practice other error processes such as photon noise may come to dominate. For a typical 4th magnitude star (H band), assuming 20\% total throughput, after 1 hour of integration the photon noise reaches a level of $\sigma = 10^{-6}$. It is also possible that some separate error process may come to dominate at a presently unknown level.

\section{Conclusions and further work}
\label{Conclusion}
The direct imaging of exo-planets is a major goal in contemporary observational astronomy, but achieving the high spatial resolutions and contrast ratios required to image solar-system scales is hampered by the Earth's turbulent atmosphere. 
Adaptive optics, interferometry, and the use of the closure-phase observable in concert has shown some success in addressing this problem. 
Pupil-remapping stellar interferometry using a monolithic photonic pupil remapper stands to vastly increase the precision of closure-phase measurement beyond the current state-of-the-art, allowing far more sensitive observations of exoplanetary systems.

This technology has been demonstrated on sky, but was previously so limited in terms of precision and throughput as to be uncompetitive. 
The root cause of these limitations was stray light propagating, unguided, from the input to the output and causing interference fringes at the output face. 

Here we presented a new generation of pupil remapper chips which overcome these limitations. 
Three key advantages have been verified.
Firstly, closure phase precision much better than one degree was demonstrated, with $\sim$0.2$^\circ$ obtainable in a single set of measurements. 
This scales to a negligible error in typical astronomical observing periods, at which point other error sources (imperfect PSF calibration, photon noise) become dominant. 
Secondly, this performance is now completely reproducible, with minimal sensitive dependence on alignment. 
Thirdly, waveguide throughputs are greatly improved, with average values of $\sim 70\%$. 
Re-routing of the waveguides addresses the unguided stray light problem, but in turn presents the problem of low throughputs due to the smaller bend radii required. 
This is then addressed with the introduction of thermally annealed waveguides.

With these fundamental problems solved, the Dragonfly instrument is now set to perform the first science observations on-sky. 
Furthermore, development can now be focused on the next evolutionary stages of the instrument. 
The use of a lithographic photonic beam combiner, instead of a free-space Fizeau type beam combiner, is being explored. 
New chips that extend operational reach beyond the current near-IR ($\sim 1.6 \mu$m) wavelengths into the mid-IR ($\sim 4 \mu$m), where the contrast ratio between a star and thermal emission from its planet is more favourable, are also being developed. 
Finally the creation of a new photonic back-end which turns Dragonfly into a nulling interferometer \cite{Labadie2007}, wherein the stellar light is interferometrically nulled to remove photon-noise from the planetary signal, is also undergoing development. 
These technologies stand to form the early steps in the photonic reformulation of astronomical imaging.

\section*{Acknowledgements}
This research was supported by the Australian
Research Council Centre of Excellence for Ultrahighbandwidth
Devices for Optical Systems (project no. CE110001018)
and the OptoFab node of the Australian National Fabrication
Facility

\end{document}